\documentclass[aps,pre,twocolumn,groupedaddress,longbibliography,floats,final,superscriptaddress]{revtex4-1}
\usepackage{xcolor}
\usepackage{graphicx}
\usepackage{amsmath}
\usepackage{amsthm}
\usepackage{bm}
\usepackage[citecolor=blue,linkcolor=blue,colorlinks]{hyperref}
\usepackage{threeparttable}
	
\definecolor{blue}{rgb}{0,0,1}
\definecolor{red}{rgb}{1,0,0}
\definecolor{green}{rgb}{0,1,0}

\newcommand{\scrH}{{\mathcal H}}
\newcommand{\scrG}{{\mathcal G}}
\newcommand{\scrI}{{\mathcal I}}

\newcommand{\scrM}{{\mathcal M}}

\newcommand{\scrR}{{\mathcal R}}
\newcommand{\scrC}{{\mathcal C}}

\newcommand{\scrW}{{\mathcal W}}

\newcommand{\bmi}{{\bm i}}
\newcommand{\bmj}{{\bm j}}
\newcommand{\bmij}{{\bm i\bm j}}

\newcommand{\KB}{K_{\rm BKT}}
\newcommand{\Kp}{K_{\rm perc}}

\graphicspath{{figures/}}

\begin{document}
\title{Percolation of the two-dimensional XY model in the flow representation}
\author{Bao-Zong Wang}
\thanks{These two authors contributed equally to this paper.}
\author{Pengcheng Hou}
\thanks{These two authors contributed equally to this paper.}
\affiliation{Hefei National Laboratory for Physical Sciences at Microscale, Department of Modern Physics, University of Science
 and Technology of China, Hefei 230027, China}
\author{Chun-Jiong Huang}
\email[]{chunjiong.huang@gmail.com}
\affiliation{Department of Physics and HKU-UCAS Joint Institute for Theoretical and Computational Physics at Hong Kong, The University of Hong Kong, Hong Kong, China}
\affiliation{Hefei National Laboratory for Physical Sciences at Microscale, Department of Modern Physics, University of Science
 and Technology of China, Hefei 230027, China}

\author{Youjin Deng}
\email[]{yjdeng@ustc.edu.cn}
\affiliation{Hefei National Laboratory for Physical Sciences at Microscale, Department of Modern Physics, University of Science
and Technology of China, Hefei 230027, China}
\affiliation{CAS Center for Excellence and Synergetic Innovation Center in Quantum Information and Quantum Physics, University of Science and Technology of China, Hefei, Anhui 230026, China}
 
\begin{abstract}
We simulate the two-dimensional XY model in the flow representation by a worm-type algorithm,
up to linear system size $L=4096$, and study the geometric properties of the flow configurations.
As the coupling strength $K$ increases, we observe that the system undergoes a percolation transition $K_{\rm perc}$ 
from a disordered phase consisting of small  clusters into an ordered phase containing a giant percolating cluster. 
Namely, in the low-temperature phase, there exhibits a long-ranged order 
regarding the flow connectivity, in contrast to the qusi-long-range order associated with spin properties. 
Near  $K_{\rm perc}$, the scaling behavior of geometric observables is well described by 
the standard finite-size scaling ansatz for a second-order phase transition. 
The estimated percolation threshold $K_{\rm perc}=1.105 \, 3(4)$ is close to but obviously smaller than 
the Berezinskii-Kosterlitz-Thouless (BKT) transition point $K_{\rm BKT} = 1.119 \, 3(10)$, 
which is determined from the magnetic susceptibility and the superfluid density.
Various interesting questions arise from these unconventional observations, and their solutions would shed lights on a variety of classical 
and quantum systems of BKT phase transitions.

\end{abstract}
\maketitle
 
\section{\label{Sec:Introduction}Introduction}

The superfluidity was first discovered in liquid helium with frictionless flow, and then became an important 
subject of persistent experimental and theoretical investigations.
In three dimensional (3D) systems, a normal-superfluid phase transitions is known to be a second-order transition 
accompanied by a Bose-Einstein condensation (BEC) with the spontaneously breaking of a $U(1)$ symmetry.
In 2D, the spontaneous-breaking continuous symmetry is forbidden by the Mermin-Wagner-Hohenberg theorem 
and BEC cannot exist. Nevertheless, the superfluidity is still developed through the celebrated Berezinskii-Kosterlitz-Thouless (BKT)  transition~\cite{berezinsky1972destruction,kosterlitz1972long,kosterlitz1973ordering,kosterlitz1974critical} 
at a finite temperature, illustrating that BEC is not an essential ingredient for superfluidity. 

In statistical mechanics, the 2D XY model is the simplest system of the normal-superfluid phase transition belonging to the BKT universality class. 
In the XY model, the superfluid density can be calculated from
 the helicity modulus (the spin stiffness) in the spin representation~\cite{Fisher1973helicity} or the mean-squared winding number in the flow representation~\cite{Pollock1987} which is similar to the case of the Bose-Hubbard model. In 2D systems, the superfluid density has a sudden jump from zero to a universal value 
 at the BKT point~\cite{Nelson1977} and this property has been used to numerically determine the BKT point~\cite{Weber1988,Schultka1994,Olsson1995,hasenbusch2005two,hasenbusch2008binder,Komura2012,Hsieh2013}. 
 Besides, the magnetic susceptibility is divergent at the BKT point as well as in the whole superfluid phase,
 which is referred as the critical region.
By the renormalization group (RG) analysis~\cite{kosterlitz1973ordering}, the correlation length exponentially diverges 
when the BKT point is approached from the disordered phase. 
For finite system sizes, this exponential divergency introduces logarithmic corrections around the BKT point,
and dramatically increases the difficulty for high-precision determination of the BKT point by numerical means 
because of the need of large system sizes and sophisticate finite-size scaling (FSS) terms.
Even though, recent Monte Carlo (MC) simulations can provide precise estimates for the coupling strength $\KB=1.119\,96(6)$~\cite{PhysRevE.65.026702,hasenbusch2005two,Komura2012}, in agreement with the high-temperature expansions~\cite{PhysRevE.79.011107}. 
It is nevertheless noted that these estimates depend on assumptions about the logarithmic finite-size corrections, 
and different extrapolations can lead to somewhat different values of the BKT point.
For instance, it was estimated $\KB= 1.119\,2(1)$ in Ref.~\cite{Hsieh2013}, which deviates from $\KB=1.119\,96(6)$ 
by about seven standard error bar.

For many statistical-mechanical systems, much insight can be gained by exploring geometric properties of the systems~\cite{arguin2002homology,morin2009critical,LIU2012107,blanchard2014wrapping,Hu2015,Hou2019,Huang2020,Newman2000,Martins2003,Hu2012,Wang2013,Xu2014,ZIFF199917,langlands1992universality,Pinson1994,PhysRevE.78.031136}.
For the Ising and Potts model, geometric clusters in the Fortuin-Kasteleyn bond representation have a percolation threshold 
coinciding with the thermodynamic phase transition,
and exhibit rich fractal properties, some of which have no thermodynamic correspondence. 
Similar behavior is observed for the quantum transverse-field Ising model in the path-integral representation~\cite{Huang2020}.
For the 2D XY model, various attempts have also been carried out. 
In Ref.~\cite{Wang2010}, geometric clusters are constructed as collections of spins 
in which the orientations of neighboring spins differ less than a certain angle.
The percolation transitions are found to be in the standard 2D percolation universality, regardless of
the coupling strength.
In Ref.~\cite{Hu2011}, spins are projected onto a random orientation, and geometric clusters 
are constructed by a Swendsen-Wang-like algorithm with an auxiliary variable. 
In the low-temperature phase $K> \KB$, a line of percolation transitions,  consistent with the BKT universality, is observed. 

In this work, we study the 2D XY model on the square lattice in the flow representation,
in which each bond between neighboring sites is occupied by an integer flow, 
and on each site, the flows obey the Kirchhoff conservation law. 
The XY model in the flow representation can be efficiently simulated 
by worm-type algorithms~\cite{Prokof2001worm,Wanwan2019}. 
Further, the superfluid density can be calculated through the winding number
and the magnetic properties can be easily measured.
From FSS analysis of the superfluid density and the magnetic susceptibility, we determine the coupling strength at the BKT transition as $\KB=1.119\,3(10)$, consistent with the most precise result $\KB=1.119\,96(6)$~\cite{Komura2012}.

Given a flow configuration, we construct geometric clusters as sets of sites connected through non-zero flows, irrespective of flow directions.
The emergence of superfluidity, having non-zero winding number, 
requests that there exists at least a percolating flow cluster.
To explore percolation in these geometric clusters, we sample the mean size of the largest clusters per site $c_1$,
which acts as the order parameter for percolation.
A percolation threshold $\Kp$ is observed. For $K < \Kp$, there are only small flow clusters, 
and $c_1$ quickly drops to zero as the linear system size $L$ increases.
For $K > \Kp$, $c_1$ rapidly converges to a $K$-dependent non-zero value, suggesting the emergence of a giant cluster and thus of 
a long-range order.
In words, as the coupling strength $K$ is enhanced, the 2D XY model in the flow configuration undergoes a percolation transition 
from a disordered phase consisting of only small clusters into an ordered phase containing a giant percolating cluster.
This is dramatically different from the magnetic properties of the 2D XY model, 
for which the system develops a quasi-long-range-order (QLRO) phase, without breaking the U(1) symmetry,
 through the BKT phase transition. 
 
The behavior of $c_1$ as a function of $K$ is very similar to the order parameter for a second-order  transition. 
To further verify this surprising observation regarding the flow connectivity, we sample the wrapping probability $R$, 
which is known to be very powerful in the study of continuous phase transitions.
It is observed that $R$ quickly approaches to 0 and 1 for $K < \Kp$ and $K > \Kp$, respectively,
and thus has a jump from 0 to 1 at $\Kp$ in the thermodynamic limit.
Near the percolation threshold, the $R$ values for different sizes have approximately common intersections, 
which rapidly converges to $\Kp$. Thus, the behavior of both $c_1$ and $R$ implies that the percolation transition is of second order.

Moreover, we find that, near $\Kp$, the FSS behavior of $R$ is well described by the standard FSS theory for a second-order transition. 
From the FSS analysis of $R$, we determine the percolation threshold $\Kp=1.105\,3(4)$ and the thermal renormalization exponent $y_t = 0.39(1)$.
The threshold $\Kp$ is close to but clearly smaller than $\KB=1.119\,96(6)$.
In addition, the estimated exponent $y_t=0.39(1)$ is significantly larger than zero.
From the FSS behavior of $c_1$, we also obtain the magnetic renormalization exponent $y_h=1.76(2)$. 
It is interesting to observe that the critical exponents ($y_t = 0.39(1), y_h =1.76(2)$) 
are not equal to $(y_t=3/4, y_h=91/48)$ for the standard 2D percolation.
These unconventional observations for the 2D XY model are much different from those for the 3D XY model, where the percolation transition and the normal-superfluid transition nicely coincide and both are the second-order phase transition~\cite{Wanwan2019}.
 
The remainder of this paper is organized as follows. Section~\ref{Sec:Model} introduces the XY model and the flow representation. 
Section~\ref{Sec:Simulation} describes the worm algorithm and sampled quantities. In Sec.~\ref{Sec:Results}, 
the MC data are analyzed and the results are presented. A brief discussion is given in Sec.~\ref{Sec:Discussion}.
 
\section{\label{Sec:Model}XY model and the flow representation}

The XY model is formulated in terms of two-dimensional, unit-length vectors
$\vec{s}\equiv(\cos{\theta},\sin{\theta})$, residing on sites of a lattice.
The reduced Hamiltonian of the XY model (already divided by
$k_B T$ , with $k_B$ the Boltzmann constant and $T$ the temperature)
reads as
\begin{equation}
\scrH = -K\sum_{\langle\bmi\bmj\rangle} \vec{s}_\bmi\cdot \vec{s}_{\bmj}
=-K\sum_{\langle\bmi\bmj\rangle}\cos(\theta_\bmi-\theta_{\bmj}),
\label{Eq:H}
\end{equation}
where the sum is over all nearest-neighbor pairs.
The partition function of Eq.~\eqref{Eq:H} can be formulated as
\begin{eqnarray}
	Z &=& \sum_{\{\vec{s}_{\bmi}\}} \exp(-\scrH)
	= \int_0^{2\pi} \prod_\bmi \frac{{\rm d} \theta_\bmi}{2\pi} \prod_{\langle\bmij\rangle} \exp\left(K \cos(\theta_\bmi - \theta_{\bmj})\right) \nonumber \\
	&=& \int_0^{2\pi} \prod_\bmi \frac{{\rm d} \theta_\bmi}{2\pi} \prod_{\langle\bmij\rangle} \left(\sum_{J_{\bmij}=-\infty}^{+\infty} I_{J_{\bmij}}(K) \exp(iJ_{\bmij}(\theta_\bmi - \theta_{\bmj}))\right) \nonumber \\
	&=& \sum_{\{J_{\bmij}\}} \prod_{\langle\bmij\rangle} I_{J_{\bmij}}(K) \int_0^{2\pi} \prod_\bmi \frac{{\rm d} \theta_\bmi}{2\pi} \exp(i\theta_\bmi \nabla \cdot \bm{\mathcal{J}}_i) \nonumber \\
	&=& \sum_{\{J_{CP}\}} \prod_{\langle\bmij\rangle} I_{J_{\bmij}}(K) \,,
\label{Eq:Z}
\end{eqnarray}
where $J_{\bmij} \in (-\infty, \infty)$ is the integer flow living on the lattice bond $\bmij$ with $J_{\bmij}=-J_{\bmj\bmi}$ and the identity $\exp(K\cos(\theta))=\sum_{J=-\infty}^{+\infty} I_J(K)\exp(iJ\theta)$ is used and $I_{J}(K)$ is the modified Bessel function. On each site $\bmi$,
$\nabla \cdot \bm{\mathcal{J}}_i=\sum_{\bmj} J_{\bmij}$ is the divergence of the flows. 
After the spin variables are integrated out, only those flow configurations, obeying the Kirchhoff conservation law
 $\nabla \cdot \bm{\mathcal{J}}_i=0$ on each site, have non-zero statistical weights. 
These flow configurations can be regarded to consist of a set of closed paths.
Figure~\ref{Fig:conf} (a) shows an example of such configurations on the square lattice with periodic boundary conditions.
 
\section{\label{Sec:Simulation}Simulation}

\subsection{Monte Carlo algorithms}
\begin{figure}[!t]
    \includegraphics[width=0.97\linewidth]{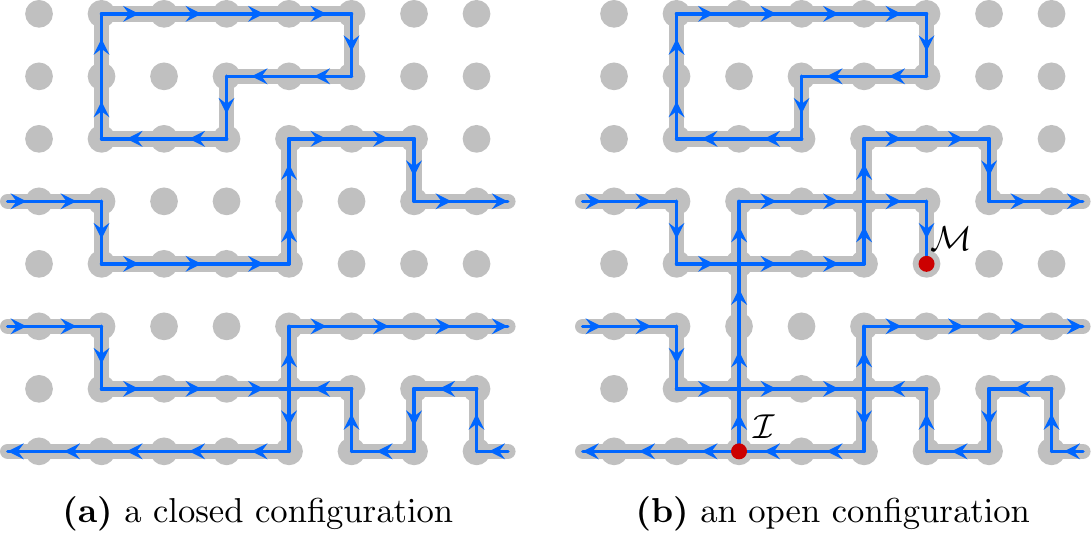}
	\caption{\label{Fig:conf} Sketch of two types of flow configurations on a square lattice with periodic boundaries. 
	Gray solid circles are the lattice sites. 
	The blue lines with arrows represent that there is a flow on a bond. 
	(a) a closed configuration with three clusters. The cluster, on the top, is local with winding number $\scrW_x=0, \scrW_y=0$ 
	and none of the wrapping event $\scrR^{(\alpha)}=0$. 
	The middle cluster wraps around the $x$ direction and has $\scrW_x\neq 0, \scrR^{(x)}=1$. 
	For the bottom cluster, although it also wraps around the $x$ direction, winding numbers are zero. 
	(b) an open configuration with two defects (red-solid circles) which do not satisfy the Kirchhoff conservation law.}
\end{figure}

The worm algorithm~\cite{Prokof2001worm} is highly efficient for loop- or flow-type representations and is employed to simulate the XY model in this work.
For worm-type simulations, the configuration space is extended to include the partition function space (the Z space) 
and a correlation function space (the $G$ space).
The $G$ space can be expressed in the flow representation as well:
\begin{eqnarray}
G &=& \sum_{\scrI\neq\scrM} \scrG(\scrI,\scrM) = \sum_{\scrI\neq\scrM} \vec{s}_{\scrI}\cdot\vec{s}_{\scrM} \exp(-\scrH) \nonumber \\
&=& \sum_{\scrI\neq\scrM} \int_0^{2\pi} \frac{{\rm d}\theta_\scrI\, {\rm d}\theta_\scrM}{(2\pi)^2} \int_0^{2\pi} \prod_\bmi \frac{{\rm d} \theta_\bmi}{2\pi}
\cosh(i(\theta_\scrI-\theta_\scrM)) \nonumber \\
&&\times
\prod_{\langle\bmij\rangle} \exp\left(K \cos(\theta_\bmi - \theta_{\bmj})\right) \nonumber \\
&=& \sum_{\scrI\neq\scrM} \int_0^{2\pi} \frac{{\rm d}\theta_\scrI\, {\rm d}\theta_\scrM}{(2\pi)^2} \int_0^{2\pi} \prod_\bmi \frac{{\rm d} \theta_\bmi}{2\pi} \cosh(i(\theta_\scrI-\theta_\scrM)) \nonumber \\
&&\times
\prod_{\langle\bmij\rangle} \left(\sum_{J_{\bmij}=-\infty}^{+\infty} I_{J_{\bmij}}(K) \exp(iJ_{\bmij}(\theta_\bmi - \theta_{\bmj}))\right) \nonumber \nonumber \\
&=& \sum_{\Delta\theta=\pm 1} \sum_{\{J_{\bmij}\}} \prod_{\langle\bmij\rangle} I_{J_{\bmij}}(K) \int_0^{2\pi}\!\!\!\! \prod_{\bmi\neq \scrI,\scrM} \frac{{\rm d} \theta_\bmi}{2\pi} \exp(i\theta_\bmi \nabla \cdot \bm{\mathcal{J}}_i) \nonumber \\
&&\times \frac{1}{2}
\int_0^{2\pi} \frac{{\rm d} \theta_\scrI\, {\rm d} \theta_\scrM}{(2\pi)^2} \exp\left(i\theta_\scrI (\nabla \cdot \bm{\mathcal{J}}_\scrI+\Delta\theta)\right) \nonumber \\
&& \times
\exp\left(i\theta_\scrM (\nabla \cdot \bm{\mathcal{J}}_\scrM-\Delta\theta)\right) \nonumber \\
&=& \sum_{\{J_{OP}\}} \prod_{\langle\bmij\rangle} I_{J_{\bmij}}(K) \,.
\label{Eq:G}
\end{eqnarray}
These none-zero weighted configurations $\{J_{OP}\}$ are called as open configurations,
in which two defects on different sites $\scrI$ and $\scrM$ are connected via an open path.
An example is shown in Fig.~\ref{Fig:conf}~(b).
In the last line of the above equation, there is no $\frac{1}{2}$ because of the exchange symmetry $\scrG(\scrI,\scrM)=\scrG(\scrM,\scrI)$.
The flow configurations in the G space also obey the Kirchhoff conservation laws for each site except $\scrI$ and $\scrM$ where $\nabla\cdot \bm{\mathcal{J}}_{\scrI}=- \Delta\theta$ and $\nabla\cdot \bm{\mathcal{J}}_{\scrM}=\Delta\theta$, with $\Delta \theta=\pm 1$.  This means that there is an additional flow $+1$ from $\scrI$ to $\scrM$ or vice versa.

The partition function in the extended configuration space is
\begin{equation}\label{Eq:Zext}
  Z_{\rm ext} = \omega_G Z + G \; ,
\end{equation}
where the relative weight $\omega_G$ between the G and the Z space can be arbitrary. 
For the particular choice $\omega_G=L^d$, the overall partition function 
becomes $Z_{\rm ext}=\sum_{\scrI,\scrM} \mathcal{G}(\scrI,\scrM) = \sum_{\{\vec{s}_\bmi\}}\left[(\sum_\bmi \vec{s}_\bmi)^2 \exp(-\scrH)\right] = \chi Z$, 
 where $\chi$ is the magnetic susceptibility. 
 
 The whole configuration space is specified by the flow variables as well as the positions of the pair of sites $(\scrI,\scrM)$. The Z space corresponds to those configurations with $\scrI=\scrM$ and this space has been expanded by $\omega_G=L^d$ times 
 due to the defect pair $(\scrI,\scrM)$ locating on an arbitrary lattice site.
In this formulation, one can naturally apply the following local update scheme: 
randomly choose $\scrI$ or $\scrM$ (say $\scrI$), move it to one of its neighboring sites (say $\scrI'$), 
and update the flow variable in between such that site $\scrI'$ becomes a new defect and the conservation law is recovered on site $\scrI$.
Effectively, the defects $(\scrI, \scrM)$ experience a random walk on the lattice.
The detailed balance condition reads as
\begin{eqnarray}
    \frac{1}{2}\frac{1}{z_d} W_{\mu} \mathcal{P}_{\mu\rightarrow\nu} &=& \frac{1}{2}\frac{1}{z_d} W_{\nu} \mathcal{P}_{\nu\rightarrow\mu} \; ,
\label{Eq:update}
\end{eqnarray}
where $z_d$ is the coordination number of the lattice and factor $1/(2 z_d)$ describes the probability 
for choosing this particular update.
Statistical weights before and after the update are accounted for by $W_{\mu}, W_{\nu}$, respectively. 
Taking into Eqs.~(\ref{Eq:Z}) and (\ref{Eq:G}) and the choice $\omega_G = L^d$, 
one has the acceptance probability according to the standard Metropolis filter as
\begin{equation}
    P_{\rm accept}=\min\left[1,\frac{I_{J_{\bmij}}^\mu(K)}{I_{J_{\bmij}}^\nu(K)}\right].
    \label{Eq:ap}
\end{equation}

The worm algorithm can be simply regarded as a local Metropolis update scheme for $Z_{\rm ext}$.
The superfluid density is measured in the $Z$ space, where the two defects coincide with each other $\scrI=\scrM$.
With the choice $\omega_G = L^d$, one has $Z_{\rm ext}=\chi Z$, and 
the magnetic susceptibility $\chi$ as the ratio of $Z_{\rm ext}$ over $Z$.
In the worm simulation, $\chi$ can be simply sampled as the statistical average of steps between subsequent closed configurations.
The relative weight $\omega_G$, of course, can take other positive value and the worm algorithm is still applicable. 
But weights of closed configurations $W_{\rm CP}$ should be scaled to $\frac{\omega_G}{L^d}W_{\rm CP}$ in Eq.~\eqref{Eq:update} and the acceptance probability needs to be modified. In this case, the worm-return time is no longer the magnetic susceptibility.

\subsection{\label{Sec:SampledQ}Sampled quantities}

In the flow representation, the winding number $\mathcal{W}_\alpha$ of a closed configuration 
is defined as the number of flows along the spatial direction $\alpha$. 
It can be calculated as $\mathcal{W}_\alpha= \sum_{\bmi} J_{\bmi,\bmi+\mathbf{\hat{e}}_\alpha}$ 
with $\mathbf{\hat{e}}_\alpha$ being the basis vector of the direction $\alpha$ and sites $\bmi$ align on a line perpendicular to the direction $\alpha$. Besides, the two-point correlations can be detected in the worm process, and the magnetic susceptibility (integral of two-pint correlation) can be evaluated by the number of worm steps between subsequent hits on the Z space,  known as worm-return time $\tau_{\rm w}$.

Given a flow configuration, we construct geometric clusters as sets of sites connected via non-zero flow variables, 
irrespective of the flow direction. 
Namely, for each pair of neighboring sites,  the bond is considered to be empty (occupied) if the flow variable is zero (non-zero),
and clusters are constructed in the same way as for the bond percolation model.
For small $K$, the flow variables are mostly zeros and the clusters are  small. 
As $K$ increases, the flow clusters grow and percolate through the whole lattice via a percolation transition. 
Following the standard insight, we measure the following observables.
\begin{enumerate}
	\item The superfluid density is calculated from  the squared winding number~\cite{Pollock1987}
	\begin{equation}
	\rho_{\rm s} = \left<\scrW_x^2+\scrW_y^2\right>/2K \,,
	\label{Eq:rho}
	\end{equation}  
	where $\left< . \right>$ represents the statistical average.
	
	\item The magnetic susceptibility $\chi=\langle \tau_{\rm w}\rangle$.
	
	\item The wrapping probability
	\begin{equation}
        R = \left<\scrR \right>,
	\end{equation}
	where we set $\scrR=1$ for the event that at least one flow cluster wraps simultaneously in two or more (x, y, or diagonal) directions. 
	 In the disordered phase, the flow clusters are too small to wrap and one has $R=0$ in the $L \rightarrow \infty$ limit. 
	 In the ordered phase with a giant percolating cluster, one has $R=1$ asymptotically. 
	 At criticality, the asymptotic value of $R$ takes some nontrivial number $0< R_c<1$. 
	 The curves of $R$ as a function of $K$ intersect for different system sizes $L$, and 
	 these intersections rapidly converge to the percolation threshold $\Kp$. 
	
	\item The size $\scrC_1$ of the largest cluster. The mean size of the largest cluster per site $c_1=\left<\scrC_1\right>/L^2$.
	In percolation, $c_1$ plays a role as the order parameter.
	In the thermodynamic limit, one has $c_1=0$ in the disordered phase and $0<c_1<1$ in the ordered phase. 
	At percolation threshold $\Kp$, it scales as $c_1 \sim L^{y_h-d}$, where $y_h$ is the magnetic renormalization exponent
	and it is also equal to the fractal dimension of percolation clusters.
\end{enumerate}
 
\begin{figure}[!t]
 	\includegraphics[width=\linewidth]{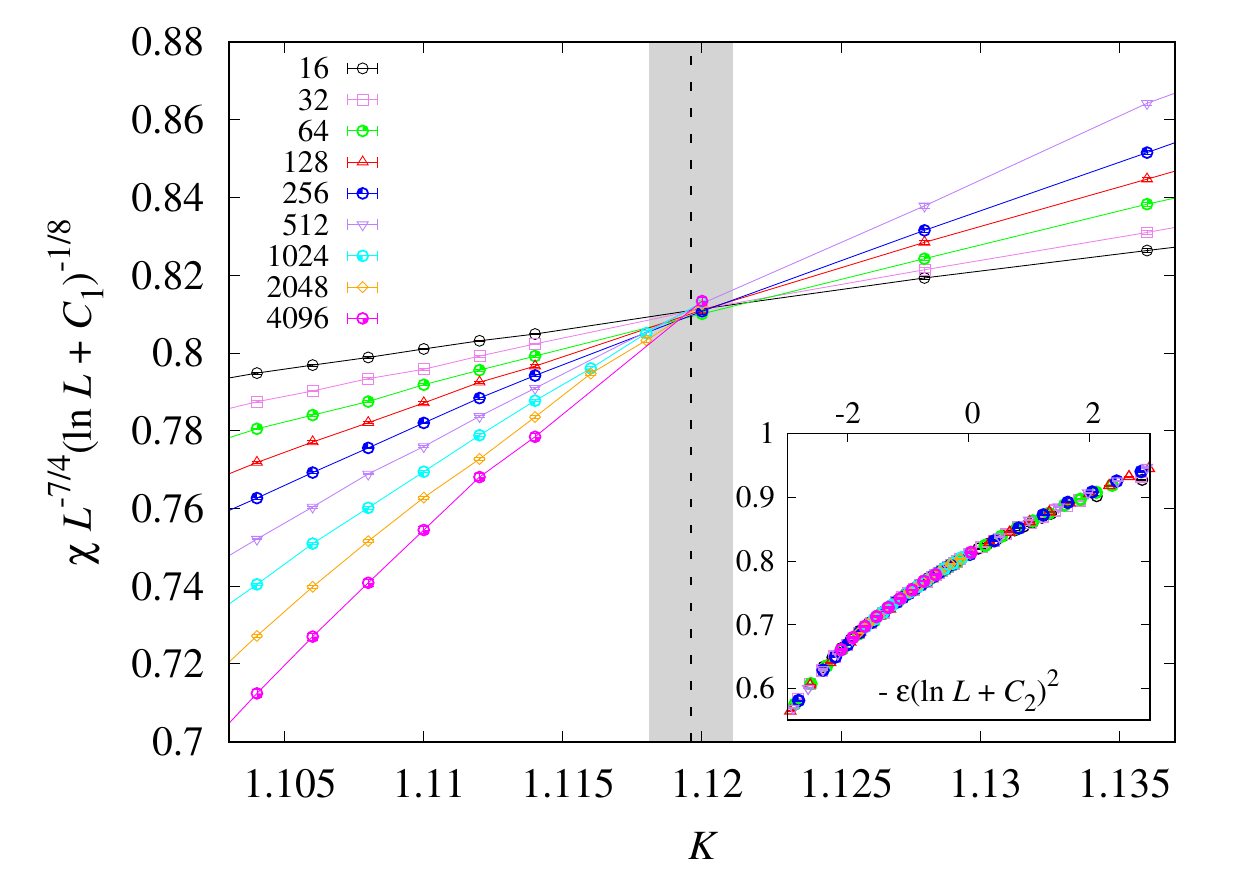}
 	\caption{\label{Fig:XYchi} Scaled magnetic susceptibility $\chi L^{-7/4}(\ln L+C_1)^{-1/8}$ 
	versus the coupling strength $K$ with $C_1=4.5$. 
	Lines  are added between the data points  for clarity. 
	The expectation value and the error bar of $\KB$ are marked with the black dashed line and the gray strip, respectively. 
	The inset shows $\chi L^{-7/4}(\ln L+C_1)^{-1/8}$ vs $-\epsilon (\ln L+C_2)^2$ with $C_2=1.57$.
	The good collapse indicates that the addictive logarithmic corrections are small,
	 consistent with the results shown in Tab.~\ref{Tab:chi}.} 
\end{figure}
 
 \section{\label{Sec:Results}Results}
 
We simulate the XY model on the square lattice with periodic boundary conditions, 
with linear system sizes in the range $16 \leq L \leq 4096$ around $K=1.11$.
After thermalizing systems to equilibration, at least $8\times 10^6$ independent samples are produced for each $K$ and $L$.

\subsection{BKT transition}

Instead of an algebraic divergence of the correlation length $\xi(t)\sim |t|^{-\nu}$ near a second-order phase transition 
with $\nu$ the correlation-length critical exponent, around the BKT transition point, the correlation length $\xi(t)$ diverges exponentially as 
\begin{equation}
	\xi(t) \sim \exp(b t^{-\frac 1 2})
	\label{Eq:xi}
\end{equation}
where $t=\KB/K-1$ is the reduced temperature and $b$ is a nonuniversal positive constant.
 This type of divergence for the correlation length leads to the logarithmic correction\cite{Weber1988,Kenna1997,Janke1997,kosterlitz1974critical}, that brings notorious difficulties for numerical study of the BKT transition. 
 Even though, in recent years, the estimates of $\KB$ have been significantly improved by 
 extensive MC simulations~\cite{hasenbusch2005two,Komura2012,Hsieh2013} and by tensor network algorithms~\cite{yu2014tensor,vanderstraeten2019approaching,jha2020critical}. 
 The most precise estimate of $\KB$ for the 2D XY model, obtained by a large-scale MC simulation with system sizes up to $L=65536$, is $\KB=1.119\,96(6)$~\cite{Komura2012}, which slightly deviates from the other MC result $\KB=1.119\,2(1)$~\cite{Hsieh2013}. 
 The complicated logarithmic corrections may be the underlying reason for the inconsistency.
 
We estimate the BKT transition point $\KB$ by studying the FSS of the magnetic susceptibility $\chi$ and the superfluid density $\rho_s$.
 \begin{table*}
	\caption{\label{Tab:chi}Fits of the magnetic susceptibility $\chi$ for the 2D XY model.}
	\scalebox{1.0}{
	\begin{tabular}{l|l|l|l|l|l|l|l|l|l}
		\hline 
		 &$L_{\rm min}$ & $\chi^2/$DF  & $\KB$  &$C_1$ &$q_0$ &$C_2$ &$q_1$ &$b_1$ &$d_1$\\
		 \hline
		 &16 & 70.4/67 & 1.119\,1(6) &4.1(10) &0.811(9) &1.44(9) &-0.057\,3(15) &0.005(12) &-0.02(7) \\
	 &32 &62.3/58 &1.119\,0(6) &4.5(9) &0.808(10) &1.54(11) &-0.055\,4(19) &0.000\,2(20) &0.01(8)  \\
		 &64 &54.2/49 &1.119\,0(11) &4.5(17) &0.808(18) &1.58(17)  &-0.055(3) &0.000\,05(329) &0.04(9)\\
		 & 64 &54.2/51 &1.118\,9(3) &4.64(22) &0.807(3) &1.57(12) &-0.055(2) &0 &0.05(9) \\
$\chi$	     &128 &37.6/42 &1.119\,2(4) &4.4(4) &0.810(5) &1.42(17) &-0.057(3) &0 &-0.18(13) \\
		 &256 &23.4/33 &1.119\,9(7) &3.5(7) &0.821(9) &0.8(3) &-0.068(6) &0 &-0.79(24) \\
		 &64 &54.5/52 &1.118\,9(3) &4.62(22) &0.807(3) &1.53(9) &-0.055\,6(16) &0 &0 \\
		 &128 &39.7/43 &1.119\,1(4) &4.5(4) &0.809(5) &1.61(12) &-0.054\,2(20) &0 &0\\
		 &256 &34.3/34 &1.119\,3(7) &4.3(7) &0.812(9) &1.60(16) &-0.054(3) &0 &0 \\
		\hline
	\end{tabular}
	}
\end{table*}
\subsubsection{Magnetic suspectibility}
According to the RG analysis, the two-point correlation function at $\KB$ scales as $G(r)\sim r^{-\eta}(\ln r)^{-2\hat{\eta}}$~\cite{kosterlitz1973ordering,kosterlitz1974critical,amit1980renormalisation}. Hence, the magnetic susceptibility $\chi$ behaves as 
\begin{equation}
\chi \sim \int_{r<\xi} d^2 r G(r) \sim \xi^{2-\eta}(\ln \xi)^{-2\hat{\eta}},
\label{Eq:chi}
\end{equation}
with the RG predictions $\eta=1/4$ and $\hat{\eta}=-1/16$. 

For finite-size systems,  it is hypothesized that the divergent correlation length near criticality is cut off by the linear system size as $\xi=\alpha L$, 
with $\alpha$ a non-universal constant. Using the linear system size, we have 
$\chi \sim L^{7/4}(\ln L+C_1)^{1/8}$, where $C_1=\ln \alpha$ is a non-universal constant. 
Together with Eq.~\eqref{Eq:xi}, one has $\alpha L\sim \exp(b t^{-1/2})$ near $\KB$, and the FSS of $\chi$ can be expressed as 
\begin{equation}
\chi(t,L) \sim L^{\frac 7 4}(\ln L +C_1)^{\frac 1 8} \widetilde \chi \left(t(\ln L+C_2)^2\right)\;,
\end{equation}
where $\widetilde{\chi}(x\equiv t(\ln L+C_2)^2)$ is an universal function and $C_2=\ln \alpha$.
Although the non-universal constants $C_1$ and $C_2$ do not affect the asymptotic scaling for $L \rightarrow \infty$,
we find that they cannot be simply neglected in finite-size analyses of MC data.

Near $\KB$, we perform least-squares fits of the $\chi$ data by the ansatz
\begin{equation}
\begin{split}
	\chi  =&  L^{\frac 7 4}(\ln L+C_1)^{\frac 1 8} \bigg[q_0 + \sum_{k=1}^{4} q_k \epsilon^k (\ln L+C_2)^{2k} \\
	 + & b_1(\ln L+C_3)^{-1}+b_2 L^{-1}+d_1 \epsilon^2 (\ln L+C_2)^2 \bigg],
\end{split}
\label{Eq:chi_fit}
\end{equation}
where $\epsilon=K_{\rm BKT}-K$, and the multiplicative and addictive logarithmic corrections have been taken into account.

As a precaution against correction-to-scaling terms that we have neglected in our chosen ansatz, we impose
a lower cutoff $L \geq L_{\rm min}$ on the data points admitted in the fit, and systematically study the effect by the chi-squared test ($\chi^2$ test) when $L_{\rm min}$ is increased. In general, our preferred fit for any given ansatz corresponds to the smallest $L_{\rm min}$ for which $\chi^2$ divided by the number of degrees of freedom (DFs) is O(1), and for which subsequent increases
in $L_{\rm min}$ do not cause $\chi^2$ to drop by much more than one unit per degree of freedom.

The results are reported in Table~\ref{Tab:chi}. In the fits with $b_1$, $b_2$ and $d_1$ free, we find that $b_1$ is consistent with zero. 
Further,  stable fits are also obtained with $b_1=b_2=0$. 
 It is worth noting that the fitting value of $d_1$ is smaller than the resolution of our fits in small $L$, but clearly nonzero when $L \geq 256$.
 This illustrates that the RG invariant function of the 2D XY model plays the role of thermal nonlinear scaling field, i.e., $a_1 \epsilon +a_2 \epsilon^2 + \cdots$~\cite{Pelissetto2013}, in which the non-universal coefficient $a_2$ cannot be neglected.

We find that the $\chi$ data for $16 \leq L \leq 4096$ and $1.104\leq K\leq 1.136$ are well described by Eq.~\eqref{Eq:chi_fit}, and we estimate the BKT transition point $\KB=1.119\,3(10)$ for the 2D XY model. Our estimate is consistent with the most precise numerical estimate $\KB=1.119\,96(6)$~\cite{Komura2012}.
The intersections, in Fig.~\ref{Fig:XYchi}, show the scaled magnetic susceptibility $\chi L^{-7/4}(\ln L+C_1)^{-1/8}$ as a function of $K$ for several system sizes. The collapse of these curves in the inset of Fig.~\ref{Fig:XYchi} confirms the scaling behavior in Eq.~\eqref{Eq:chi_fit}.

\subsubsection{Superfluid density}
\begin{figure}[!t]
    \includegraphics[width=0.85\linewidth]{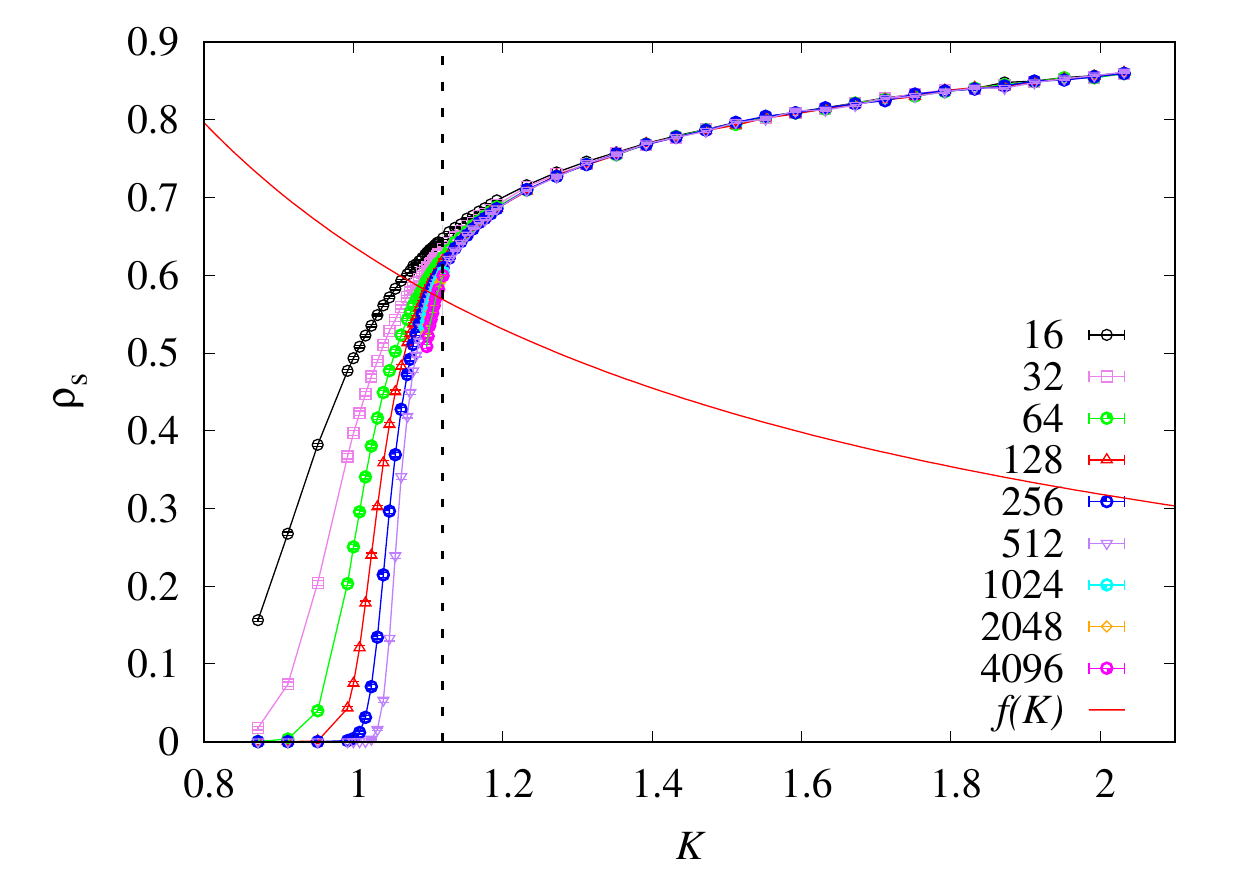}
    \caption{\label{Fig:XYrho} Superfluid density $\rho_{\rm s}$ versus the coupling strength $K$. 
    The lines connecting the data points are added for clarity. 
    The vertical dashed line is the BKT transition coupling $\KB\approx 1.1193$ (as determined in this paper) of the 2D XY model. 
    The red line is $f(K)=2f_{\rm r}/(\pi K)$, with $f_{\rm r} = 1-16\pi e^{-4\pi}$.
    The superfluid density $\rho_s(\KB,L)$ will approach to the intersection of the dash line 
    and the red line with increasing system sizes $L$.}
\end{figure}

In the renormalization-group analysis, 
the superfluid density $\rho_s^{\rm r}$ has a jump at the BKT point as the temperature decreases, 
and, according to the Nelson-Kosterlitz criterion~\cite{Nelson1977},  the size of the jump is given by 
\begin{equation}
\lim_{\begin{subarray}{l}K \to \KB^- \\ L \to \infty\end{subarray}} \rho_{\rm s}^{\rm r}(K,L)= \frac{2}{\pi \KB}.
\end{equation}
In MC study of the 2D XY model, the situation is more subtle: the size of the jump depends on how the thermodynamic limit is approached. 
More precisely speaking, the jump of the superfluid density $\rho_{\rm s}$, calculated from the mean-square winding number of Eq.~\eqref{Eq:rho},
becomes $f_{\rm r} \rho_{\rm s}^{\rm r}$, where the factor $f_{\rm r}$  depends on
the aspect ratio $L_x/L_y$,  with $L_x$ and $L_y$ being the linear sizes along the $x$ and $y$ directions, respectively.
 For the case of $L_x=L_y$, one has $f_{\rm r}=1-16\pi e^{-4\pi}$, as proved in Ref.~\cite{Prokofev2000}. 
 
 The MC data for $\rho_s$ are shown in Fig.~\ref{Fig:XYrho}, where the slow convergence of $\rho_s$ at the BKT point 
 is due to logarithmic corrections. 
 Around $\KB$, we perform least-squares fits of the $\rho_s$ data  by 
\begin{equation}
\rho_s  =  \rho_{\rm s,c} + \sum_{k=1}^{3} q_k \epsilon^k (\ln L+C)^{2k}
+b_1(\ln L+C^\prime)^{-1}+b_2 L^{-1},
\label{Eq:W_fit}
\end{equation}
where $\epsilon=\KB-K$, $\rho_{\rm s,c}=2f_{\rm r}/(\pi \KB)$, and the leading logarithmic correction has been taken into account. 
The results are summarized in Tab.~\ref{Tab:rho_s}. With $C$ and $C^\prime$ being free parameters, we have stable fits 
with $b_1$ free and $b_2=0$. We obtain $\KB=1.119\,3(10)$, consistent with our estimate from $\chi$.

Similar to $\chi$, the logarithmic corrections of $\rho_s$ exist and some literatures achieve different estimates of the BKT point by 
analyzing the FSS of $\rho_s$~\cite{Weber1988,Schultka1994,Olsson1995,hasenbusch2005two,hasenbusch2008binder,Komura2012,Hsieh2013}, because of different forms of the logarithmic corrections.

\begin{table}
	\caption{\label{Tab:rho_s}Fits of the superfluid density $\rho_s$ 
		for the 2D XY model.}
	\scalebox{0.95}{
	\begin{tabular}[t]{l|l|l|l|l|l|l|l}
			\hline 
			&$L_{\rm min}$ & $\chi^2/$DFs  & $\Kp$ &$C$ & $q_1$  &$C^\prime$ &$b_1$\\
			\hline
			&32 & 23.2/55 & 1.119\,4(4) &5.3(6) &-0.013(2) &0.25(13) &0.247(10) \\
			$\rho_s$      &64 &21.1/47 &1.119\,3(6) &5.5(7) &-0.013(2) &0.2(3) &0.234(18) \\
			&128 &18.3/39 &1.119\,2(8) &5.3(9) &-0.013(3) &0.2(5) &0.23(3) \\ 
			\hline
		\end{tabular}
	}
\end{table}

\begin{figure}
    \includegraphics[width=0.75\linewidth]{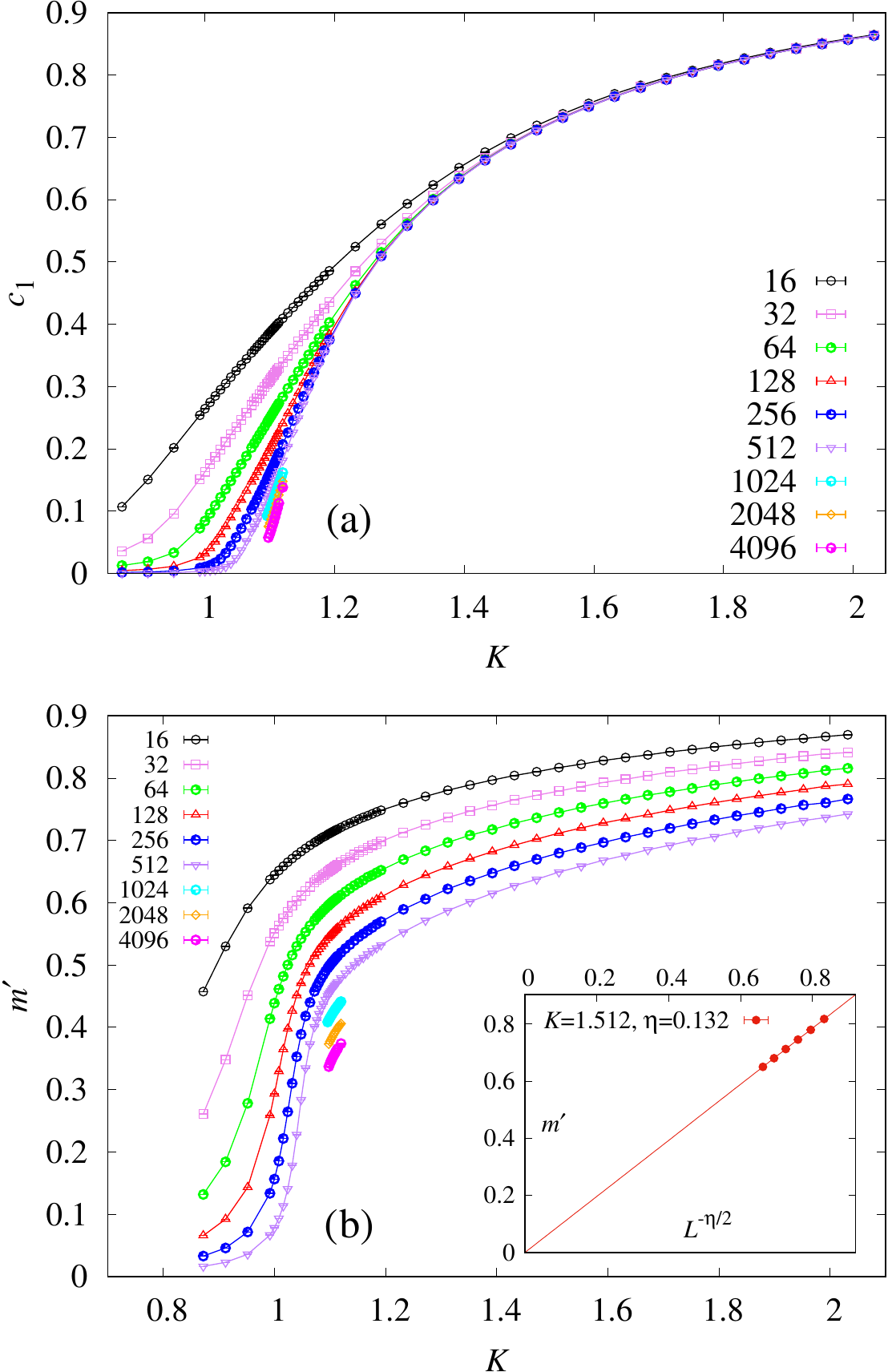} \\
    \caption{\label{Fig:rho1} Mean size of the largest cluster per site $c_1$ and 
    the magnetization-density-like quantity $m'$ versus coupling strength $K$. 
    (a) $c_1$ v.s. K. The curves of different system sizes converge to a single curve in the ordered phase. 
    (b) $m'$ v.s. $K$. In the whole low-temperature phase, due to the absence of spontaneous-symmetry breaking, 
    $m'$ decreases monotonically with increasing system size and reaches zero in the thermodynamic limit. 
    As an example, the inset illustrates 
    the algebraic decay of $m'$ as a function of $L$ for $K=1.512$.
    }
\end{figure}

\subsection{\label{Sec:pt} Geometric properties}

To have an overall picture of the geometric properties of the flow clusters, we simulate the 2D XY model with the coupling strength 
in a relatively wide range $ 0.84 < K < 2.04$. 
Figure~\ref{Fig:rho1}(a) shows the mean size of the largest cluster per site $c_1$, 
which plays a role of the order parameter for percolation.
The behavior of $c_1$ as a function of $K$  is very similar to that for a conventional percolation transition. 
In the disordered phase with small $K$, all the flow clusters are small and finite, and $c_1$ quickly drops to zero as system size $L$ increases;
in the ordered phase, a giant percolation cluster emerges and thus a long-range order develops. 
For $K \geq 1.3$, Figure~\ref{Fig:rho1}(a)  clearly shows that $c_1$ rapidly converges 
to a non-zero value.

\begin{figure}
    \includegraphics[width=0.85\linewidth]{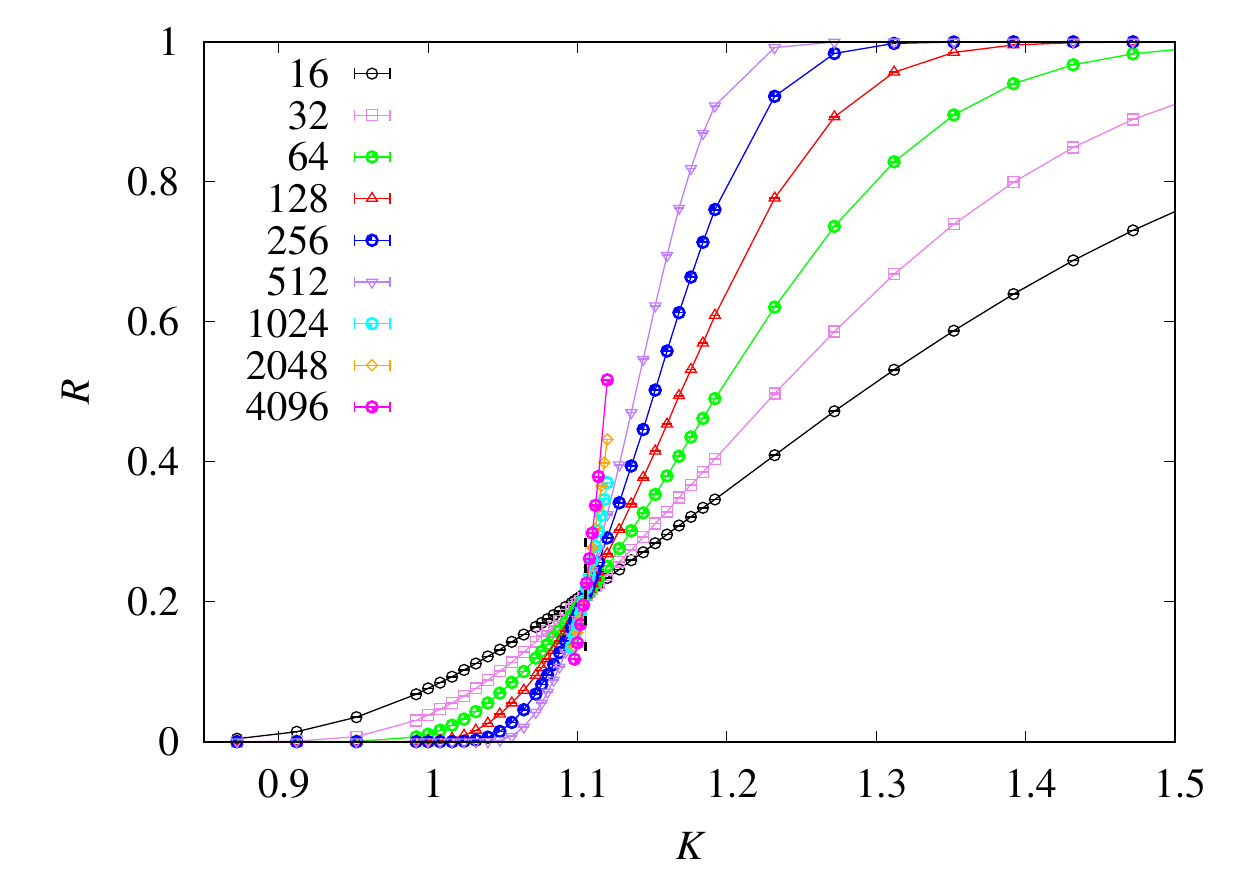}
    \caption{\label{Fig:R2} Wrapping probability $R$ versus coupling strength $K$.}
\end{figure}

Figure~\ref{Fig:rho1}(a) indicates that, for the 2D XY model, the geometric features of the flow configurations 
are very different from the spin properties. 
In the flow representation, since the spin degrees of freedom are integrated out, 
we cannot directly sample the magnetization 
density $m$--the order parameter for spin properties.
Nevertheless,  the magnetic susceptibility $\chi$ relates to  $m$ as $\chi = L^d \langle m^2 \rangle$,
and we can define an effective parameter as $m' \equiv \sqrt{\chi/L^d}$. As shown in Fig.~\ref{Fig:rho1}(b), 
$m'$ also drops rapidly to zero in the disordered phase ($K < \KB$), similar to $c_1$.
However, in the low-temperature phase ($K > \KB$), $m'$ keeps decreasing as $L$ increases,
which is still clearly seen for $K$ as large as $K \approx 2$. 
According to the RG analysis, 
the whole region for $K > \KB$ is critical, 
one has an algebraic decay $m' \sim L^{-\eta}$ for $K > \KB$, where $\eta$ is a $K$-dependent exponent.
As an illustration, the inset of Fig.~\ref{Fig:rho1}(b) displays the algebraic decay of $m'$ for $K=1.152$, with $\eta \approx 0.132$.

To further demonstrate the second-order-like percolation transition of the flow clusters, 
we plot in Fig.~\ref{Fig:R2} the wrapping probability $R$ versus $K$. 
In the absence or presence of a giant cluster, one expects $R \rightarrow 0$ or 1 in the $L \rightarrow \infty$  limit, respectively.
This is indeed supported by Fig.~\ref{Fig:R2}, in which the wrapping probability $R$ quickly converges to 1 as long as $K > 1.2$,
illustrating the emergence of a giant cluster penetrating the lattice.
Moreover, as for a second-order phase transition, the $R$ curves for different system sizes 
have an approximately common intersection, 
indicating the location of the percolation threshold $\Kp$.
As $L$ increases, the intersection of the $R$ curves quickly approaches to $\Kp$.

In short, the scaling behaviors of $c_1$ and $R$ as a function of $K$ are both consistent with those for a second-order phase transition, 
instead of a BKT transition. 
This is an unconventional and surprising phenomenon.

\subsubsection{Percolation threshold}
\begin{figure}
    \includegraphics[width=0.85\linewidth]{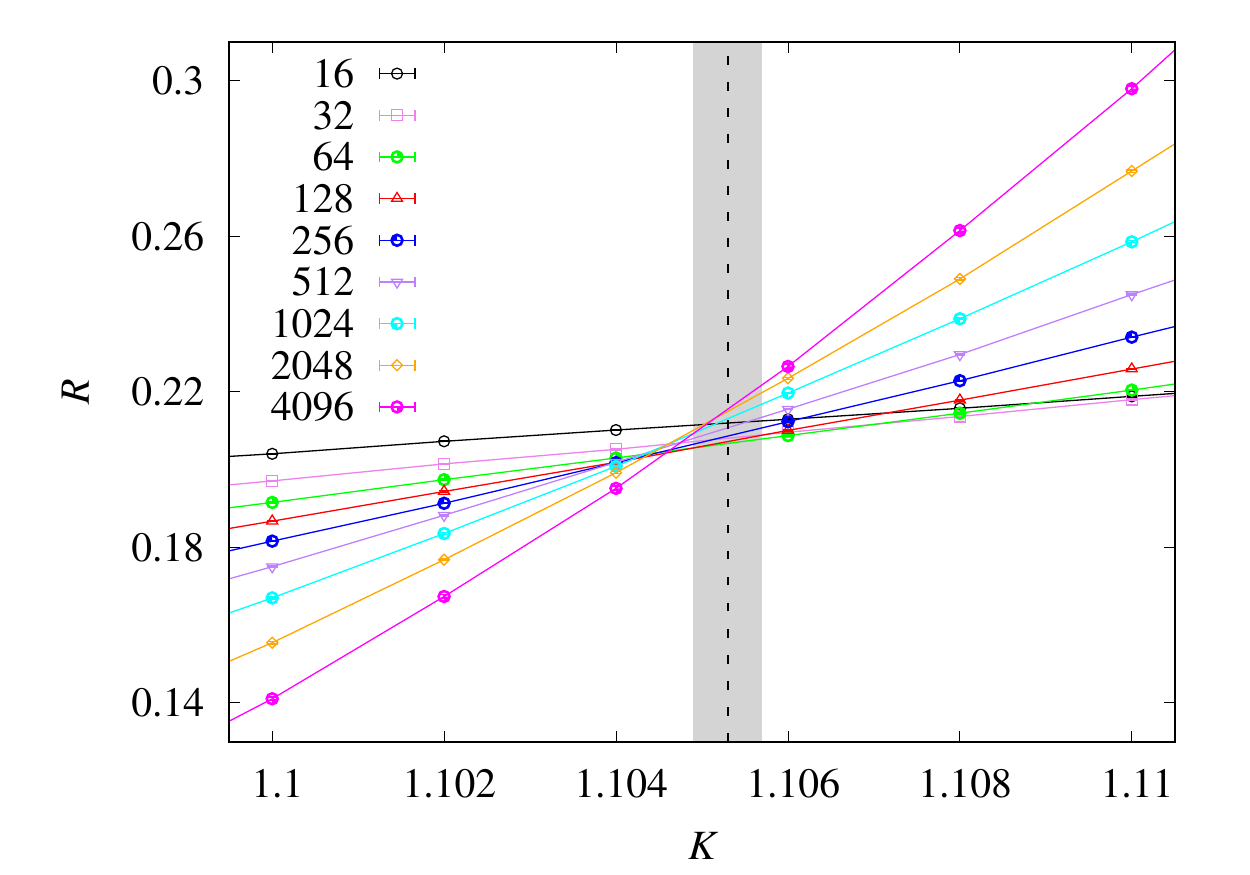} \\
    \caption{\label{Fig:XYR2} Wrapping probability $R$ versus coupling strength $K$ around $\Kp$.
    The lines connecting the data points are added for clarity. 
    The expectation value and the error bar of $\Kp$ are marked with the black dashed line and the gray strip, respectively.}
\end{figure}

To have a quantitative numerical estimate of the percolation threshold $\Kp$, 
we plot in Fig.~\ref{Fig:XYR2} the MC data for the wrapping probability $R$ near $\Kp$.
It can be seen that the uncertainty of the intersections of $R$ for sizes $ L \in [16, 4096]$ 
is at the third decimal place, varying in range $ 1.104 < K < 1.110$. 
As $L$ increases, the intersection moves downward from $K \approx 1.110$ 
for $L \approx 32$ to $K \approx 1.104$ for $L \approx 512$, 
and then slightly moves upward to $K \approx 1.105$. 

\begin{table*}
	\caption{\label{Tab:Wrapping}Fits of the wrapping probability $R$.}
	\scalebox{1.0}{
		\begin{tabular}{l|l|l|l|l|l|l|l|l|l}
			\hline 
			 &$L_{\rm min}$ & $\chi^2/$DF  & $\Kp$  &$y_t$ &$R_c$ &$q_1$ &$b_1$ &$c_1$ &$d_1$ \\
			 \hline
			&64 &60/38 & 1.104\,95(5) &0.390(3) &0.211\,2(6) &-0.62(1) &-1.9(2) &3.3(4) &-1.7(5) \\
	$R$	&128 &32.2/31 &1.105\,30(9) &0.392(4) &0.217(2) &-0.61(2) &-4.6(6) &3.6(9) &-1.9(6)  \\
			&256 &27.9/24 &1.105\,5(2) &0.393(6) &0.220(4) &-0.62(3) &-8(3) &4(2) &-1.9(7)\\
	    \hline
		\end{tabular}
	}
\end{table*}

\begin{table}[!t]
	\caption{\label{Tab:C1}Fits of the mean size of the largest cluster per site $c_1$.}
	\scalebox{1.0}{
	\begin{tabular}{l|l|l|l|l|l|l}
	\hline 
	& $L_{\rm min}$ &$\chi^2/$DF &$y_h$ &$a_0$ &$a_1$ &$y_1$\\
	\hline
	& 16 &1.7/5 &1.773(7) &0.46(5) &0.50(4) &-0.214(12) \\
$c_1$	&32 &1.4/4 &1.768(10) &0.49(7) &0.47(6) &-0.23(3)\\
	& 64 &1.0/3 &1.762(13) &0.54(10) &0.44(7) &-0.25(6)\\
	\hline
	\end{tabular}
	}
\end{table}

As in the earlier discussions, the percolation transition of the flow configurations looks like a second-order transition. 
Near $\Kp$,  the $R$ data in Fig.~\ref{Fig:XYR2} are indeed well described by the following standard FSS ansatz 
for a continuous phase transition 
\begin{equation}
    R(\epsilon,L) = \tilde{R}(\epsilon L^{y_t}) \; ,
\label{eq:R}
\end{equation}
where $\tilde{R}$ is a universal function and $y_t=1/\nu$ is the thermal renormalization exponent.
Taylor expansion of Eq.~(\ref{eq:R}) leads to
\begin{equation}
\begin{aligned}
R(\epsilon,L)=& R_c +\sum_{k=1}^3 q_k \epsilon^k L^{k y_t} + b_1 L^{y_i} + b_2 L^{y_i-1} \\
	&+ b_3 L^{y_i-2}+ c_1\epsilon L^{y_t+y_i}+d_1 \epsilon^2 L^{y_t},
\end{aligned}
\label{Eq:fit_O}
\end{equation}
where $\epsilon=K-\KB$ and the terms with exponent $y_i <0$ account for finite-size corrections.
We fit the $R$ data by Eq.~(\ref{Eq:fit_O}), and find that the correction exponent is $y_i\approx -1$. 
The results with $y_i=-1$ are summarized in Table~\ref{Tab:Wrapping}.
We obtain $\Kp=1.105\,3(4)$ and $y_t=0.39(1)$, of which the error bar of $\Kp$ is at the fourth decimal place.

Assuming that the precision of $\Kp$ is reliable,  we conclude that 
the percolation threshold $\Kp=1.105\,3(4)$ is significantly smaller than the BKT transition $\KB=1.119\,96(6)$.
Actually, the deviation, at the second decimal place, can be already seen from an bare eye view of Fig.~\ref{Fig:XYR2}.
Therefore, our numerical data suggest that the percolation threshold does not coincide with the BKT transition.
It is noted that, since the emergence of superfluidity requests the existence of a percolating cluster, 
one must have $\Kp \leq \KB$, which is indeed satisfied in our results.

The estimated thermal exponent $y_t=1/\nu=0.39(1)$ is much larger than zero. 
It it were true, the characteristic radius of the geometric clusters would diverge as a power law $\sim \epsilon^{-\nu}$,
different from the exponential growth of the correlation length near the BKT transition. 
This provides another piece of evidence that the percolation transition is not BKT-like.
In addition, since the standard uncorrelated percolation in 2D has the thermal exponent $y_t=3/4$, 
the result $y_t=0.39(1)$ suggests that the percolation of the flow configurations is not in the 2D percolation universality class.

Further we fit the data for the mean  size  of  the largest cluster per site $c_1$ at $K=1.106$ by the FSS ansatz
\begin{equation}
    c_1(L)= L^{y_h}(a_0 +a_1 L^{y_1}),
\end{equation}
where $y_h$ is the magnetic exponent. 
The results are shown in Table~\ref{Tab:C1}, and we have $y_h=1.76(2)$, smaller than $y_h=91/48 \approx 1.89$ 
for the 2D percolation universality.

\section{\label{Sec:Discussion}Discussion}

We simulate the XY model on the square lattice in the flow representation by a variant of worm algorithm. 
From the FSS analysis of the magnetic susceptibility $\chi$ and the superfluid density $\rho_s$, 
we estimate the BKT transition to be $\KB=1.119\,3(10)$, consistent with the most precise result $\KB=1.119\,96(6)$.

We study the geometric properties of the flow configurations by constructing clusters 
as sets of sites connected through non-zero flow variables.
An interesting observation is that, in the low-temperature phase, there is a giant cluster that 
occupies a non-zero fraction of the whole lattice, indicating the emergence of a long-range order parameter 
for the flow connectivity. 
Given a flow configuration, a non-zero winding number of flows implies a superfluid state,
and can occur only if at least a flow cluster wraps around the lattice. 
Such a percolating cluster can be either giant or fractal; for the latter, the cluster size per site vanishes 
in the thermodynamic limit.
Since the low-temperature phase of the 2D XY model is a quasi-long-range-ordered state, 
the flow clusters are expected to be fractal. 
The unexpected emergence of a giant cluster raises an important question:
what is the nature of the percolation transition separating the disordered phase of small clusters 
and the ordered phase of a giant cluster?

The overall behaviors of the size of the largest cluster per site $c_1$ and of the wrapping probability $R$ 
indicate that the percolation transition is of a second order. 
Further, the $R$ data near the threshold $\Kp$ are well described by a standard finite-size scaling ansatz 
for a continuous phase transition. 
From the least-squares fits of $R$, we obtain the percolation threshold as $\Kp=1.105\,3(4)$,
which is close to but clearly smaller than the BKT point $\KB=1.119\,96(6)$.
The  thermal exponent $y_t=1/\nu = 0.39(1)$ is also significantly larger than zero.
This implies an algebraic divergence of the characteristic radius of the flow clusters, 
instead of an exponential growth of the correlation length near the BKT transition. 

We determine the magnetic renormalization exponent as $y_h=1.76(2)$ from the size of the largest cluster.
The set of critical exponents $(y_t=0.39(1),y_h=1.76(2))$ significantly deviates from $(y_t=3/4,y_h=91/48)$ 
for the standard percolation in 2D. 
With the assumption that the estimated error margins are reliable, we obtain that the percolation transition 
of the flow clusters belongs to a new universality.

Many open questions arise from these unconventional observations. 
First, since the difference between $K_{\rm BKT}$ and $K_{\rm perc}$ is at the second decimal place, 
can it be simply due to complicated logarithmic FSS corrections that have not been carefully taken into account in the analyses?
If this were the case,  the intersections of $R$ for different system sizes would eventually converge to $K_{\rm BKT} \approx 1.20$. 
However, as shown in Fig.~\ref{Fig:XYR2}, the intersections of $R$ are mostly in range $K \in (1.104, 1.106)$, except for some small sizes.
Thus, finite-size corrections would change dramatically for $L > 4096$ if the final convergence is near $K \approx 1.20$.
To clarify this point, simulation for $L \gg 4096$ is needed, which is beyond our current work.
Second, what is the nature of the percolation transition for the flow clusters? 
Figures~\ref{Fig:rho1}(a) and \ref{Fig:R2} indicate that in the low-temperature region, 
a giant cluster emerges and a long-range order parameter  develops for percolation. 
Therefore, with the assumption that there is only one percolation transition, 
the disordered phase of small flow clusters and the ordered phase of a giant cluster 
are expected to be separated by a second-order transition,
 consistent with the behaviors of $c_1$ and $R$. 
Third, what universality does the percolation transition belong to, if it were of a second order?
The estimated critical exponents $(y_t=0.39(1),y_h=1.76(2))$ suggest that the percolation is not in the same universality as 
the standard percolation in 2D. Fourth, do these unconventional phenomena occur in other systems exhibiting the BKT transition?

A possible scenario is that, as the coupling strength $K$ is enhanced, the 2D XY model in the flow representation 
first experiences a second-order percolation transition $\Kp$ for the flow connectivity 
and then enters  into the superfluidity phase via the BKT transition $\KB $.
In the flow configurations, the superfluid flows for $K > \KB$ live on top of the giant cluster, which already appears when $K > \Kp$.
In the small intermediate region $\KB > K > \Kp $, the giant cluster, while wrapping around the lattice, 
is effectively built up by a set of local flow loops and thus no superfluidity occurs.  
For this scenario, a deep understanding of the  physics in the intermediate region is still needed.
For instance, does the emergence of the giant flow cluster have relations to 
the turbulent behavior of the large amount of unbound vortices immediately above the BKT transition? 

Beside the 2D XY model, there exist many systems of the BKT transition, 
and the Bose-Hubbard (BH) model is a typical example of such systems. 
Given a finite temperature, as the on-site coupling strength is decreased, the 2D BH model 
undergoes a BKT phase transition from the normal fluid into the superfluid phase.
Using a worm-type quantum Monte Carlo algorithm, we simulate the 2D BH model in the path-integral representation,
and obtain evidence that the percolation threshold of the flow clusters does not coincide with the BKT transition. 
Future works shall focus on an extensive study of low-dimensional quantum systems exhibiting the BKT phase transition.

\bibliography{bibliography.bib}
 
\end{document}